\documentstyle[aps,epsfig,preprint]{revtex}
\newcommand \ee{\end{equation}}
\newcommand \be{\begin{equation}}

\newcommand \bea {\begin{eqnarray} \nonumber }
\newcommand \eea {\end{eqnarray}}
 \def\(({\left(}
 \def\)){\right)}
\def\[[{\left[}
\def\]]{\right]}

\begin{document}

\title{Directed polymer in random media, in two dimensions~:
numerical study of the aging dynamics}
\author{A. Barrat\footnote{Present address~: International Center
for Theoretical Physics, Strada Costiera 11, 34100 Trieste, Italy~;
email barrat@ictp.trieste.it}}

\address{Laboratoire de Physique Th\'eorique de l'Ecole
Normale Sup\'erieure\footnote{Unit\'e propre du CNRS, associ\'ee \`a
 l'Ecole Normale Sup\'erieure et \`a l'Universit\'e de
Paris Sud}, 24 rue Lhomond, 75231 Paris Cedex 05, France}

\maketitle

\begin{abstract}

Following a recent work by Yoshino, we study the aging dynamics
of a directed polymer in random media, in $1+1$ dimensions.
Through temperature quench, and temperature cycling numerical
experiments similar to the experiments on real spin glasses, we
show that the observed behaviour is comparable to the one of
a well known mean f\/ield spin glass model. The observation of
various quantities (correlation function, ``clonation'' overlap
function) leads to an analysis of the phase space landscape.
\end{abstract}

\begin{center}
LPTENS preprint 96/57
\end{center}


\vskip 1cm

\section{Introduction}

The study of directed polymers in random media has triggered a lot
of interest and of works (for a recent review see \cite{revuepol}),
since it is related to many f\/ields, from the f\/luctuations
of interfaces \cite{interf} to quantum mechanical problems in
a time-dependent random potential \cite{quantum}, or the very
topical problem of vortex lines in high temperature superconductors
\cite{revuevortex}. It has also connections with spin glasses, as has
been shown by Derrida and Spohn \cite{derridaspohn}, who studied a
mean field version (on a Cayley tree) of the random polymer, showing
the existence of a low temperature phase similar to the
Random Energy Model of Derrida \cite{rem}.

In f\/inite dimensions ($d+1$, with $d$ transverse dimensions), the
existence of a phase transition has been shown for $d \ge 2$
\cite{imbrie,cook,derridapol}, whereas the system is always in a low
temperature phase for $d=1$, and has been called a ``baby spin
glass'' in \cite{marc90}.

The dynamics of such model has not been much studied so far. A
recent numerical work by Yoshino \cite{yoshino} has made clear
the existence of aging for the directed polymer in random media
in dimensions $1+1$, with violation of time translation invariance
and of the f\/luctuation-dissipation theorem. The observed behaviour
for the correlation function is similar to the scaling properties
for the simulations of a three dimensional spin glass model \cite{rieger}.
It is analyzed along the lines of a scenario similar to the
droplets model for spin glasses \cite{fisherhuse1,fisherhuse2}~:
the polymer moves in a network of ``tubes'' where its probability
of presence (calculated at equilibrium with a transfer matrix
method) is high. The network has a quite complicated spatial
structure, and the tubes form loops of various sizes. The dynamics
consists then of rapid f\/luctuations inside the tubes (acting as
traps) combined with thermally activated jumps between different
tubes. These thermal excitations are compared to the droplets
excitations. Besides, the fact that the loops display a broad
distribution of sizes \cite{hwafisher} induces a broad distribution of
relaxation times, and thus aging.

On the other hand, Cugliandolo, Kurchan and Le Doussal, following
the study of the aging dynamics of a mean-f\/ield spin glass
model \cite{cukuprl}, and of
a particle in an inf\/inite dimensional random potential
\cite{franzmezard,cuglledou}, have proposed an analytical
treatment of the long times off equilibrium dynamics of an elastic
manifold embedded in an inf\/inite dimensional space, in the presence
of a quenched random potential \cite{cukuledou} (the statics of
such a model has been studied by M\'ezard and Parisi with
a replica variational Gaussian approximation, becoming exact
in this inf\/inite dimensional limit \cite{mezpar91}). This
corresponds here to an inf\/inite $d$. The f\/inite dimension of
the manifold leads to the study of the relaxation of its Fourier
modes $k$. The two-times correlation and response functions
$C_k(t,t')$ and $r_k(t,t')$ satisfy dynamical equations where
two regimes can be separated, as for the case of the particle
\cite{franzmezard,cuglledou})~: a stationary regime, and a regime
displaying aging, where the properties of equilibrium dynamics
(namely, time translation invariance and f\/luctuation dissipation
theorem) are violated. Various equations can be written for
the long times behaviour of the functions, for example, the
measure of the violation of the f\/luctuation dissipation
theorem $X(C)$ does not depend on $k$.

We will here focus on the same model as \cite{yoshino}~; after
describing this model, the used dynamics, we perform
several numerical experiments for aging dynamics. Preparing the
polymer in its ground state, we also show that it can display
stationary dynamics, and compare the two kinds of dynamics
in order to analyse the phase space landscape.

\subsection{Model}

The polymer is def\/ined on a square lattice of linear size $N$~: it
consists of $N$ monomers lying on the sites
$\{ (i,x_i)$, $i=1, \cdots , N \}$, and the Hamiltonian is~:
\be
H [\{x_i\}] = \sum_{i=2}^{N} \((\vert x_i - x_{i-1} \vert + V(i,x_i)
\)),
\ee
where $V$ is a Gaussian random potential, with zero mean and
variance $\sigma$, uncorrelated from site to site.
The f\/irst term gives the elastic energy~; moreover, the steps
$\vert x_i - x_{i-1} \vert$ are restricted to $0$ or $1$,
and one of the extremities is f\/ixed, so that a Boltzmann measure
can be def\/ined~: $x_1=N/2$.

Transfer matrix method can be used to study the statics
of this model \cite{marc90}. In particular, the Edwards-Anderson
parameter can be calculated, following \cite{marc90}~: we take
two replicas of the polymer, with the same realization
of the disordered potential~; if
$Z(x,y,L)$ is the partition function for the pairs of
polymers arriving at transverse coordinates $x$ and $y$ after
$L$ steps (or monomers), and if
$\tilde{Z}(x,y,q,L)$ is the partition function restricted to such
pairs having an overlap $q$, it is possible to write recursion
relations in $L$ for $Z$ and
$Y(x,y,L)= \sum_{q} q\ \tilde{Z}(x,y,q,L)$, and therefore to
evaluate
\be
\lim_{L \to \infty} \frac{1}{L}
\frac{\sum_{x,y} Y(x,y,L)}{\sum_{x,y} Z(x,y,L)} ,
\ee
and to average it over disorder to obtain $q_{EA}$. We will use
this value to check some expected long time limits of
dynamical quantities (see below).

The directed polymer is evolving with Monte-Carlo dynamics
in a heat bath in the following way~: a monomer and a move are
chosen at random, and the move is performed 
with probability $\min(1,\exp(-\beta \Delta E))$ (Metropolis
algorithm), $\beta$ being the inverse temperature, and
$\Delta E$ the change of energy involved. One Monte-Carlo step
consists in $N$ such tries.

\subsection{Numerical experiments}

The transfer matrix method allows us to f\/ind the ground state in
a given realization of a potential~; we have studied
the dynamics in two cases~: the initial conf\/iguration is either
taken at random, or as the ground state. The polymer, therefore
coming from an inf\/inite temperature or from a zero temperature
thermalized state, is then free to evolve at the temperature of the
heat bath. We then measure the evolution of the energy, the two-times
correlation function def\/ined by
\be
C(t_w+t,t_w) = \overline{\langle
\sum_{i=1}^N \delta_{x_i(t_w+t) , x_i(t_w)} \rangle },
\ee
where $\langle . \rangle$ is a mean over thermal noise
and the overline denotes a mean over realizations of the quenched
disorder,
and the overlap between two copies of the polymer evolving
in the same realization of the potential, with the same thermal
noise until $t_w$, and then decoupled
\cite{andrea1,andrea2,cugldean,babumezov} (a process called
``clonation'' in \cite{andrea1})~;
if these copies are labeled by $^{(1)}$ et $^{(2)}$, this
overlap is
\be
Q_{t_w} (t_w+t,t_w+t) =  \overline{\langle
\sum_{i=1}^N \delta_{x_i^{(1)}(t_w+t) ,  
x_i^{(2)}(t_w+t)} \rangle }.
\ee
Besides, we have performed temperature cycling experiments in the
same way as for real spin glasses \cite{saclay}.
Most of the runs have been made with a polymer of length
$N=500$, and some with $N=800$~: no f\/inite size ef\/fects were
seen for the used simulation times.

\section{Aging dynamics}

\subsection{Quench at initial time}

\begin{figure}
\centerline{\epsfig{figure=figure1a.eps,width=8.5cm,angle=-90}}
\centerline{\epsfig{figure=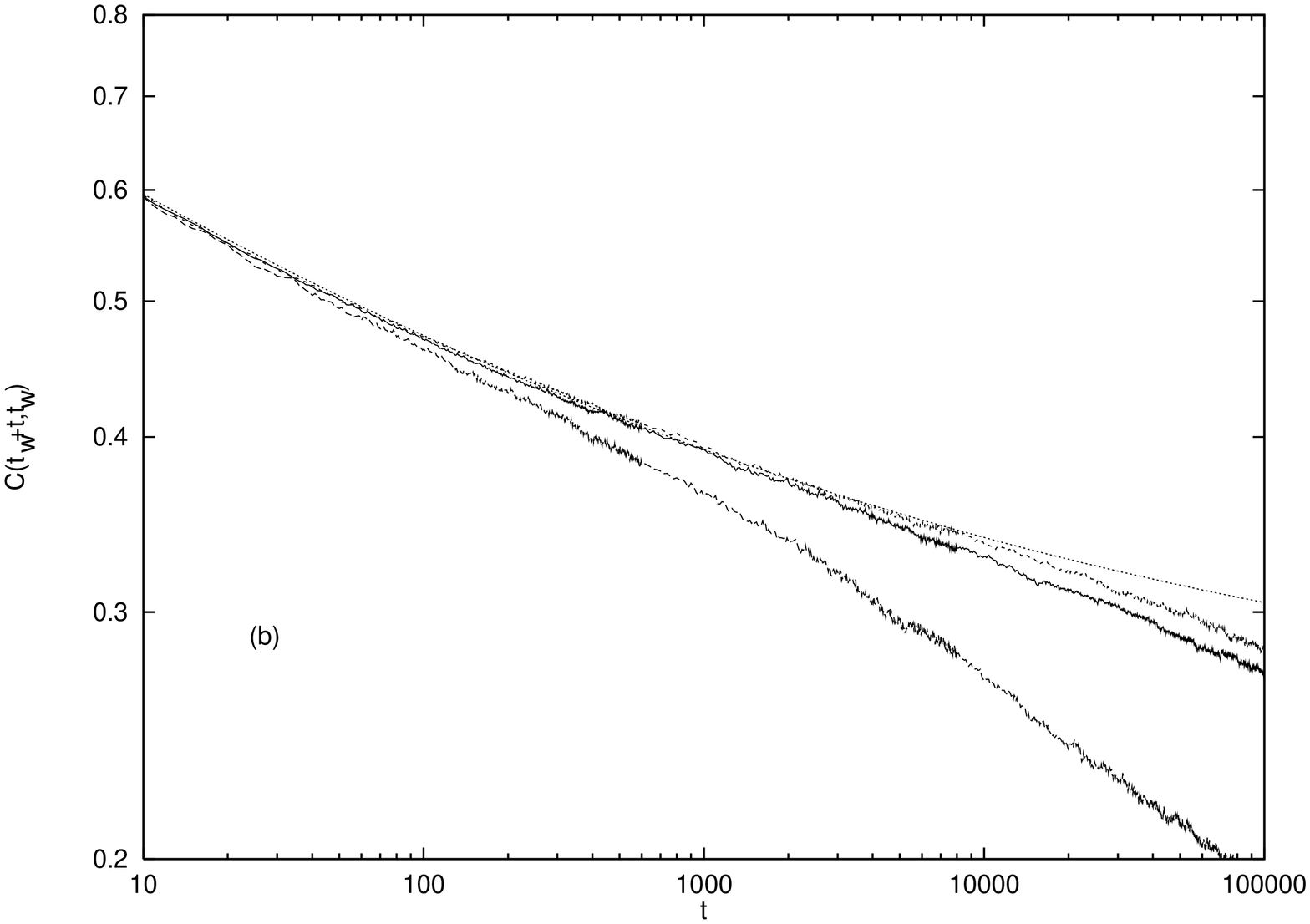,width=8.5cm,angle=-90}}
\caption{(a)~: from bottom to top,
$C(t_w+t,t_w)$ versus $t$ for
$t_w=50,\ 500,\ 5000,\ 100000$, at $T=1$, and
the fit $C_{as}(t)=0.65 + .25\ t^{-0.08}$. (b)~:
$C(t_w+t,t_w)$ versus $t$ for $T=3$, and $t_w = 2000,\ 50000,\
100000$, and $C_{as}(t)=0.24 + 0.545\ t^{-0.185}$. In both cases,
$q_{EA}(T=1)=0.65$ and $q_{EA}(T=3)=0.24$ have
been obtained by transfer matrix method.
}
\label{correlation}
\end{figure}

The initial conf\/iguration of the polymer is chosen at random,
with the constraint $\vert x_i - x_{i-1} \vert = 0\  \mbox{or}\ 1$~; then
it evolves at a f\/ixed temperature. It has already been observed
\cite{yoshino} that the correlation function
$C(t_w+t,t_w)$ previously def\/ined displays aging behaviour~: it
depends explicitly on $t_w$ and $t$ (see f\/igure (\ref{correlation}))~;
as the system ages, it becomes more rigid, in the sense that it evolves
slower and gets away from himself always slower.

For times $t$ much lower than $t_w$, the dynamics has the characteristics
of equilibrium dynamics~: $C(t_w+t,t_w)$ depends only on $t$ (time
translation invariance), and it has been checked numerically
\cite{yoshino} that the f\/luctuation-dissipation theorem is valid. Besides,
we have also checked the validity of the relation
$Q_{t_w}(t_w+t,t_w+t)= C(t_w+2t,t_w)$ \cite{babumezov}.
In this regime, we have therefore a quasi-equilibrium dynamics. Some
well-known spin glass models (like for example the $p$-spin
spherical model \cite{cukuprl}, or the random manifold
\cite{franzmezard,cukuledou}) present a correlation function
decaying from $1$ to $q_{EA}$, with a power law
approach to the $q_{EA}$ plateau~: $q_{EA}+ A t^{-\nu}$.

\begin{figure}
\centerline{\epsfig{figure=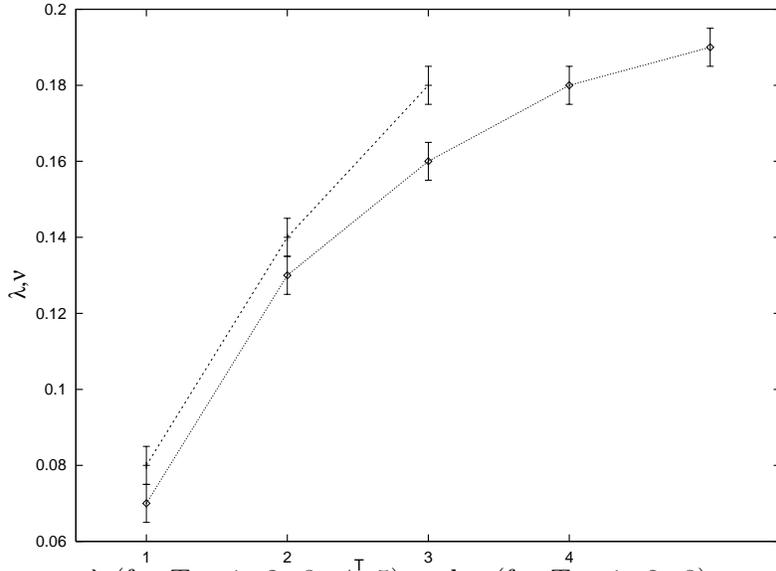,width=7.5cm,angle=-90}}
\caption{Exponents $\lambda$ (for $T=1,\ 2,\ 3,\ 4,\ 5$)
 and $\nu$ (for $T=1,\ 2,\ 3$) versus temperature.}
\label{lambda}
\end{figure}

\begin{figure}
\centerline{\epsfig{figure=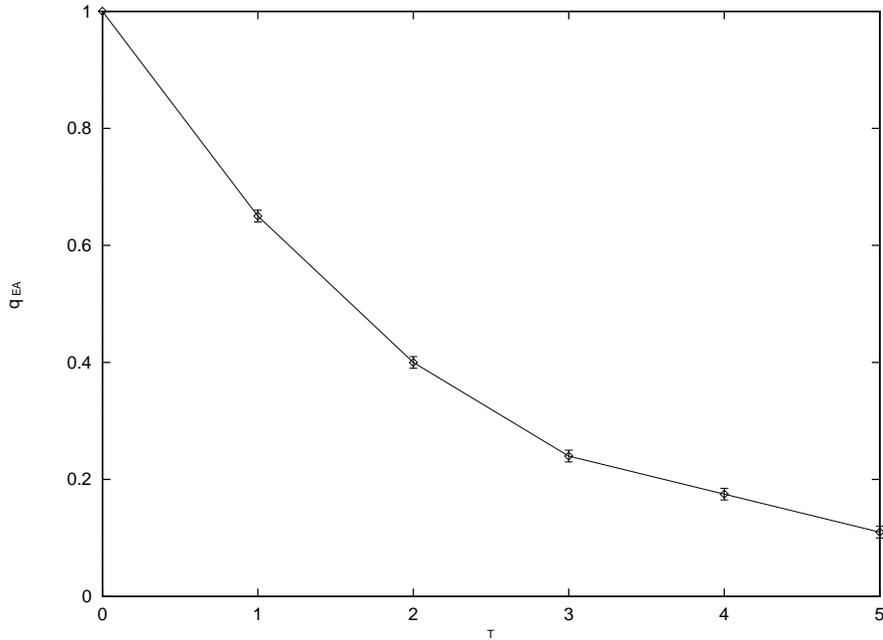,width=8.5cm,angle=-90}}
\caption{
Values of $q_{EA}$ obtained by transfer matrix method,
versus temperature.
}
\label{qea}
\end{figure}

We have been able to f\/it the results of the simulations with
such form, using the values of $q_{EA}$ obtained by transfer matrix
method~: this fits therefore use only $2$ parameters,
and not $3$ (see f\/igure (\ref{qea}) for the values, and
f\/igure (\ref{correlation}) for examples of the f\/its).
The obtained values of $\nu$ (typically in the range
$0.1 - 0.2$) are displayed in f\/igure
(\ref{lambda}).

For times $t$ comparable or bigger than $t_w$, the correlation function
$C(t_w+t,t_w)$ depends explicitly on $t_w$~:
the dynamics is no more time translation invariant. For $t \gg t_w$,
it decays as a power law,
\be
C(t_w+t,t_w) \simeq f(t_w,T) (t/t_w)^{-\lambda}.
\ee
We are therefore in presence of a weak-ergodicity breaking
behaviour, $\lim_{t \to \infty} C(t_w+t,t_w) =0$ \cite{bouchaud}.
The obtained values of $\lambda$ are comparable to the values
of $\nu$ (see f\/igure (\ref{lambda})), and both exponents are
increasing functions of temperature, like in real spin glasses
\cite{saclay}.

It is important to note that the $\nu$ exponent is dif\/ferent
from the $x$ exponent studied by Yoshino \cite{yoshino}~:
this $x$ is obtained by the scaling form
\be
C(t_w+t,t_w) \simeq  t^{-x}
\ee
for the $t \ll t_w$ part, with a global form
\be
C(t_w+t,t_w)=  t^{-x} \Phi (t/t_w).
\ee
If we take the limit $\lim_{t_w \to \infty} \lim_{t \to \infty}$ we obtain
the same behaviour, but the opposite order of limits,
$\lim_{t \to \infty} \lim_{t_w \to \infty} C(t_w+t,t_w)$ yields $0$,
in contradiction with the expected static limit
\be
\lim_{t \to \infty} \lim_{t_w \to \infty} C(t_w+t,t_w)=q_{EA},
\ee
which is also obtained by the form $q_{EA}+ A t^{-\nu}$.

With numerical data it is however
dif\/f\/icult to prefer one of these forms, and
much longer simulations (much bigger values of $t_w$) would be necessary.

\begin{figure}
\centerline{\epsfig{figure=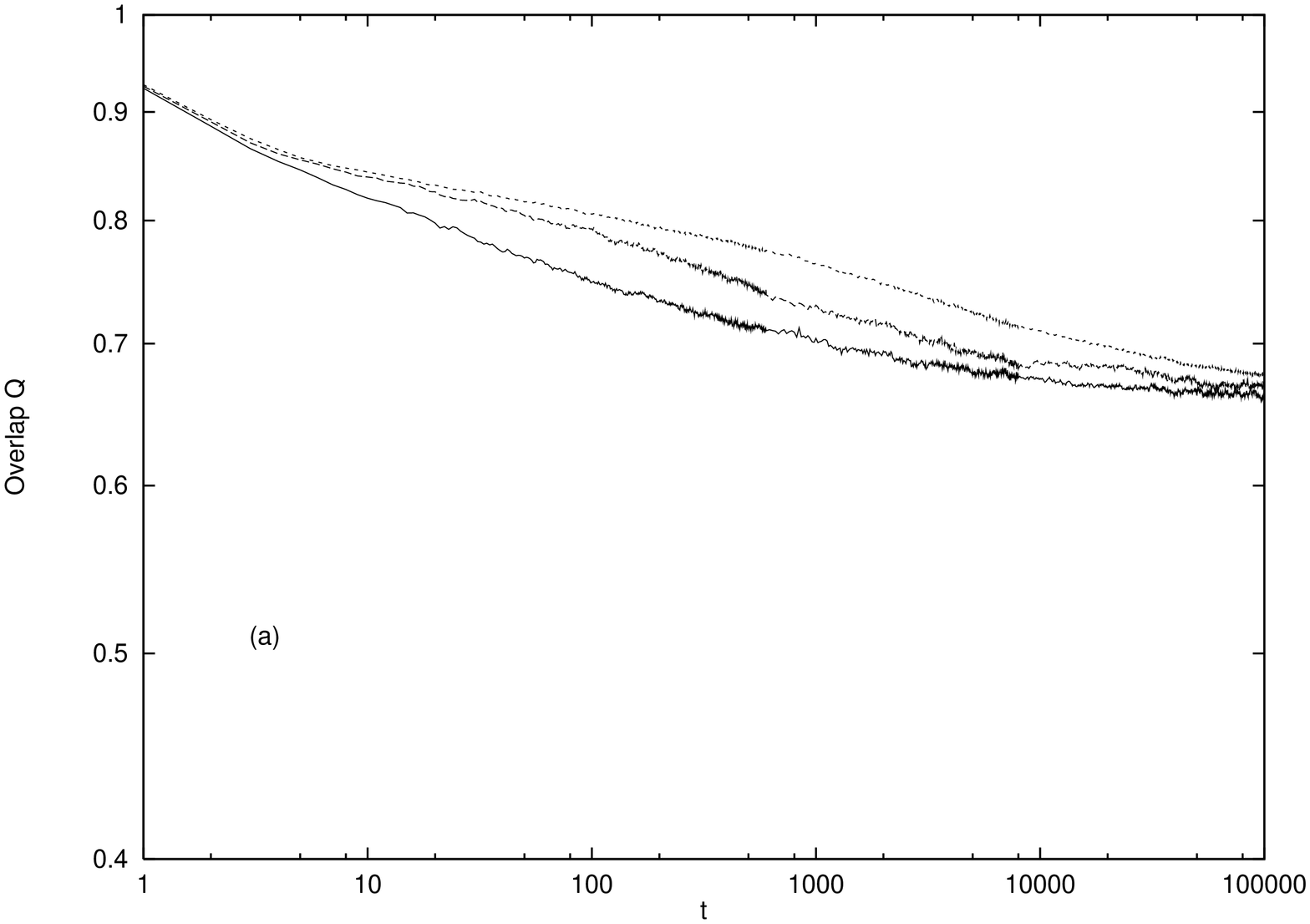,width=8.5cm,angle=-90}}
\centerline{\epsfig{figure=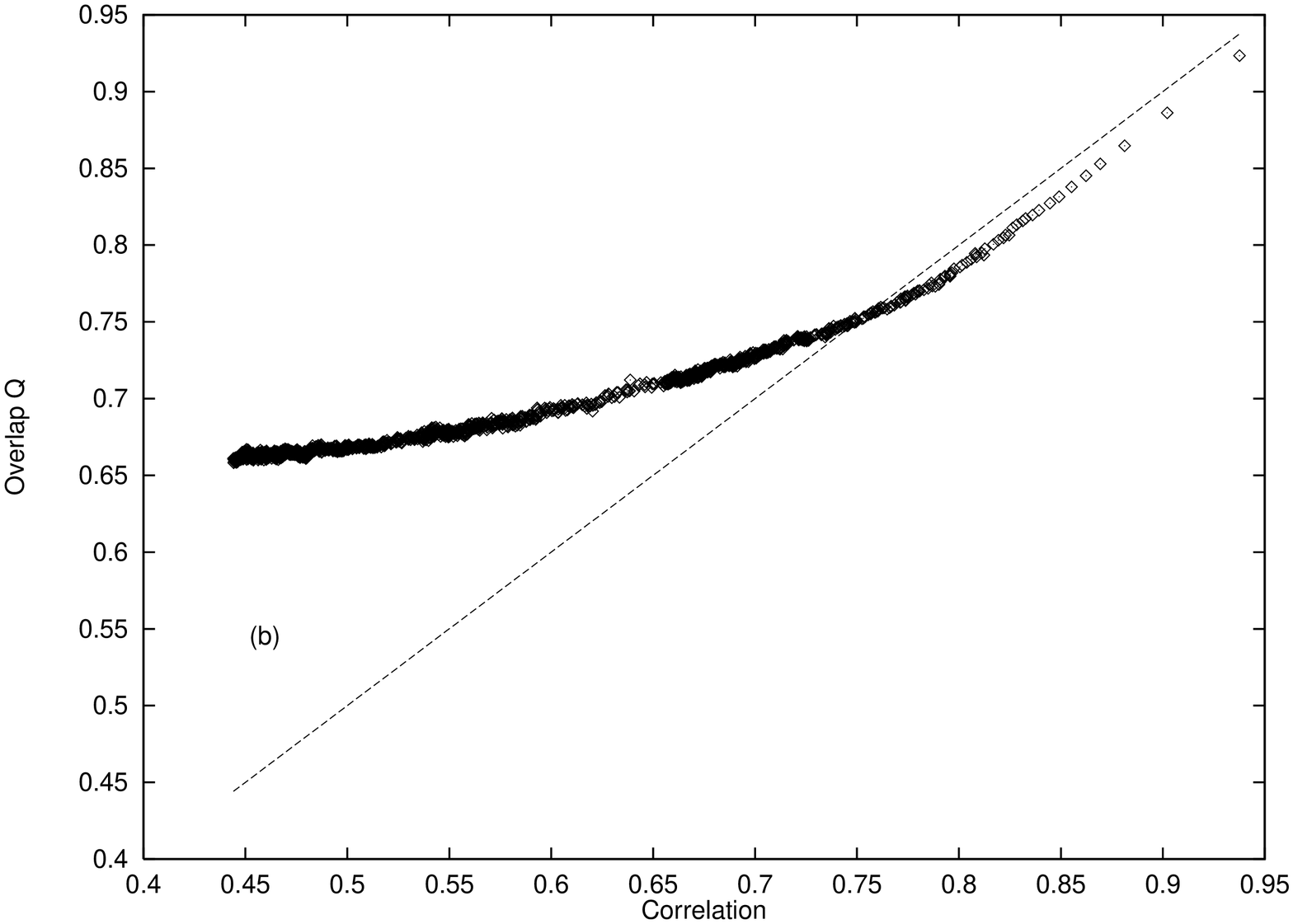,width=8.5cm,angle=-90}}
\caption{$Q_{t_w}(t_w+t,t_w+t)$ for $t_w=50,\ 500,\ 5000$, $T =1$,
in logarithmic scale (a)~; $Q_{t_w}(t_w+t,t_w+t)$ versus
$C(t_w+t,t_w)$ for $t_w=50$, $T =1$
(b). At $T=1$, $q_{EA} \approx 0.65$.
}
\label{overlap}
\end{figure}

If we now look at the overlap between two replicas separated at $t_w$,
$Q_{t_w}$, it seems that this function has a f\/inite limit
at large times $t$ (see f\/igure (\ref{overlap})), with a value
compatible with the value of $q_{EA}$, for big enough $t_w$.
This constatation puts this model in the class I of
the classif\/ication of \cite{babumezov}, which includes
domain-growth models, and the $p=2$ spherical $p$-spin
model \cite{cugldean}~: it indicates that the evolution
in phase space takes place in ``corridors'' \cite{andrea1}, of size
$q_{EA}$. On the contrary, the case of the manifold embedded
in a infinite dimensional space yields
$\lim_{t_w \to \infty} \lim_{t\to \infty} Q_{t_w}(t_w+t,t_w+t)=0$, and
therefore belongs to type II, which
probably indicates a much more complex phase-space landscape, and occurs
also for example for the $p$-spin spherical model with $p \ge 3$
\cite{babumezov}.

\subsection{Temperature cycling experiments}

\begin{figure}
\centerline{\epsfig{figure=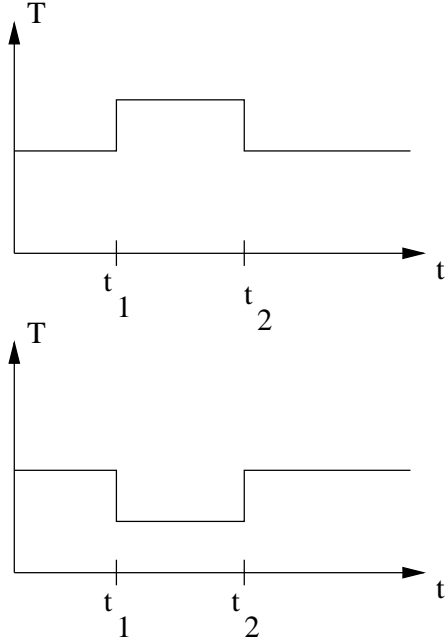,width=8.5cm,angle=-90}}
\caption{Temperature cycles.}
\label{sauts1}
\end{figure}

A spin-glass quenched under its transition temperature, and then submitted
to temperature changes, shows a very puzzling behaviour
(see \cite{saclay} for a review, and references therein) that we
brief\/ly describe now.
After a quench at time $t=0$,
the temperature cycle is as follows (see f\/igure (\ref{sauts1}))~:
the temperature is $T$ from $t=0$ to
$t=t_1$, then $T + \Delta T$ from $t=t_1$ to $t=t_2$, and again
$T$ after $t=t_2$~; 
$\Delta T$ can be negative as well as positive.
For a positive $\Delta T$, one observes a reinitialization of
the dynamics, with, e.g. for a thermoremanent magnetization, a
relaxation  after $t_2$
identical to the one obtained after a quench at $t=0$
and a waiting time $t_{eff}=t_w-t_2$. On the contrary, for a negative
$\Delta T$, the system keeps memory of its evolution, and its relaxation
corresponds to an ef\/fective age between $t_w$ and $t_w-t_2+t_1$.

We have performed these jumps numerically with $T=2$,
$\Delta T= 1,\ 2,\ -1$, and $t_1=500$, $t_2=1500$.
We have then monitored the evolution
of the energy of the polymer, as well as $C_{jump}(t_w+t,t_w)$
with $t_w=2000$, comparing these quantities with the
ones obtained for $\Delta T= 0$ (constant temperature).

\begin{figure}
\centerline{ \epsfig{figure=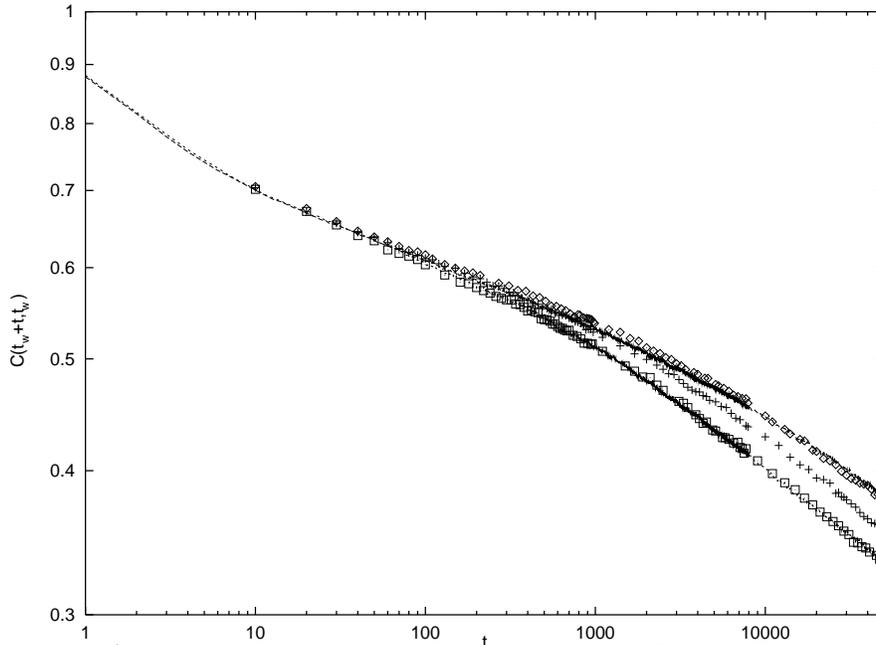,width=8.5cm,angle=-90}}
\caption{Results of the temperature cycling experiments~:
The symbols show $C(t_w+t,t_w)$ at constant temperature $T=2$
for $t_w=1000$ (squares), $t_w=2000$ (crosses),
$t_w=3000$ (diamonds). 
The temperature cycles are done with $t_1=500$, $t_2=1500$,
$t_w=2000$, $T=2$, and yield $C_{jump}(t_w+t,t_w)$~:
the corresponding curves
are superimposed on $C(1000+t,1000)$ for $\Delta T=-1$,
and on $C(3000+t,3000)$ for $\Delta T=1$.}
\label{sauts}
\end{figure}

It is clear (f\/igure (\ref{sauts})) that
$C_{jump}(t_w+t,t_w)$ corresponds to a certain
$C(t_{eff}+t,t_{eff})$, $t_{eff}$ being an
ef\/fective age for the system, depending on $t_2 - t_1$ and on
$\Delta T$~; $t_{eff}$ is less than $t_w$ for a negative $\Delta T$, which
means that
the dynamic has been slowed down by the time spent at a lower
temperature~; here we estimate $t_{eff}=1000 = t_w -t_2 +t_1$
(for $\Delta T=-1$, the data corresponding to
$C_{jump}(2000+t,2000)$ with $t_1=500$, $t_2=1500$ are superimposed
on the curve $C(1000+t,1000)$). For smaller
values of $\Delta T$, or longer times $t_2-t_1$, $t_{eff}$ can be bigger
than $t_w -t_2 +t_1$~: the time spent at $T+\Delta T$
can contribute a little to the aging.
For positive values of $\Delta T$, $t_{eff}$ is bigger than
$t_w$ (in f\/igure (\ref{sauts}), $t_{eff}=3000$),
showing that the time spent at $T+\Delta T$ has contributed to the
approach to equilibrium more than the same time at $T$.

The behaviour of the directed polymer is therefore symmetrical
for positive or negative variations of temperature. No reinitialization
of dynamics is found. This type of behaviour is similar to the one
observed for mean-f\/ield spin glasses \cite{saclay,cugldean}, and thus
very dif\/ferent from the one observed in real experiments on spin glasses.

It should be remarked that real experiments deal with response
function, whereas we are monitoring correlation functions. However,
numerical simulations of a three dimensional Ising spin glass model
have shown also for the correlation functions
\cite{riegercycling} a partial reinitialization of the
dynamics for positive $\Delta T$, and asymmetric outcomes of
numerical experiments with positive or negative temperature cycles. Nothing
of this kind is found here.

\section{Relaxation from the ground state}

The transfer matrix method allows to f\/ind the ground state of the
polymer, given a realization of the potential. We then let the polymer
evolve at temperature $T$ as before.

\begin{figure}
\centerline{\epsfig{figure=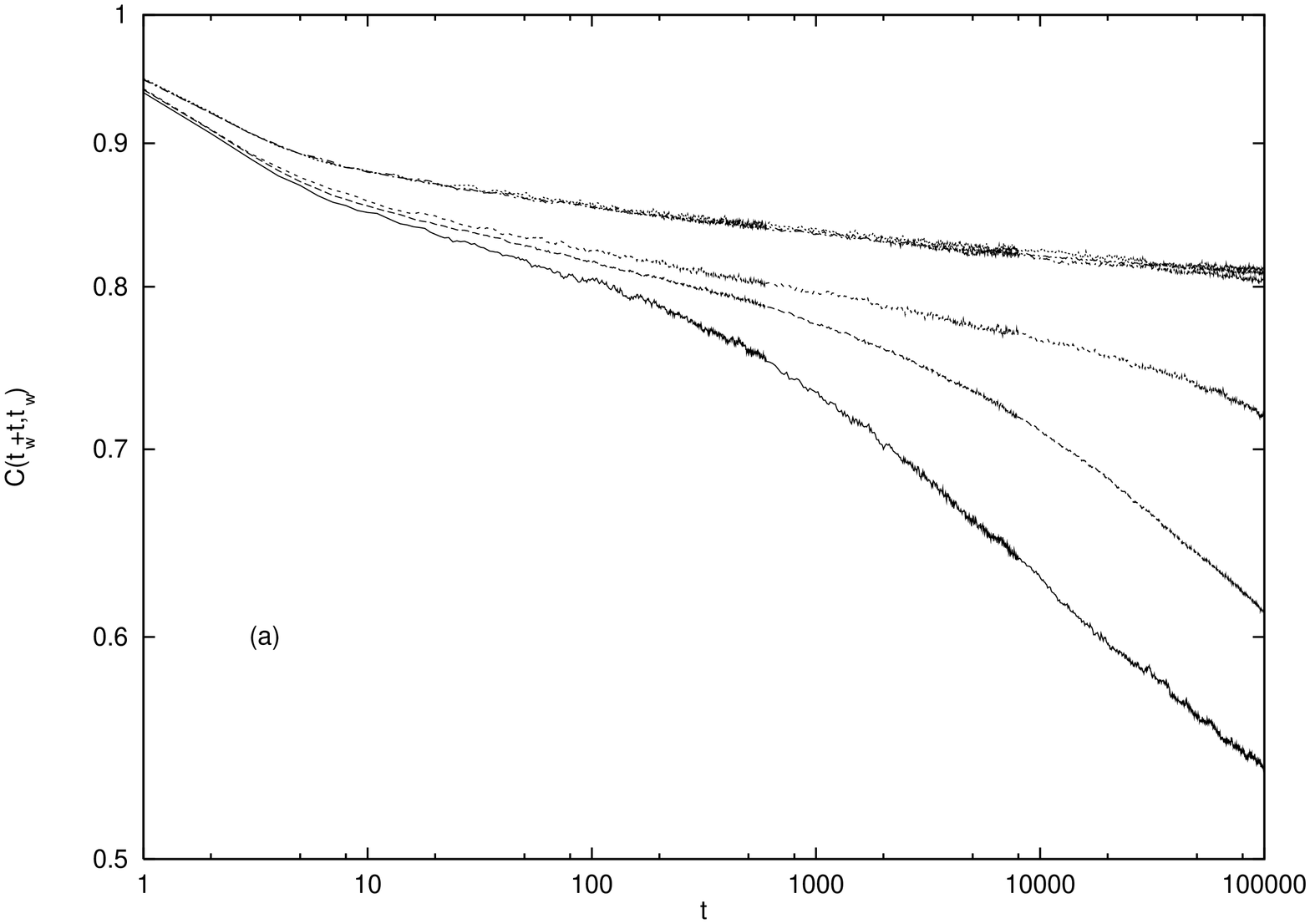,width=8.5cm,angle=-90}}
\centerline{\epsfig{figure=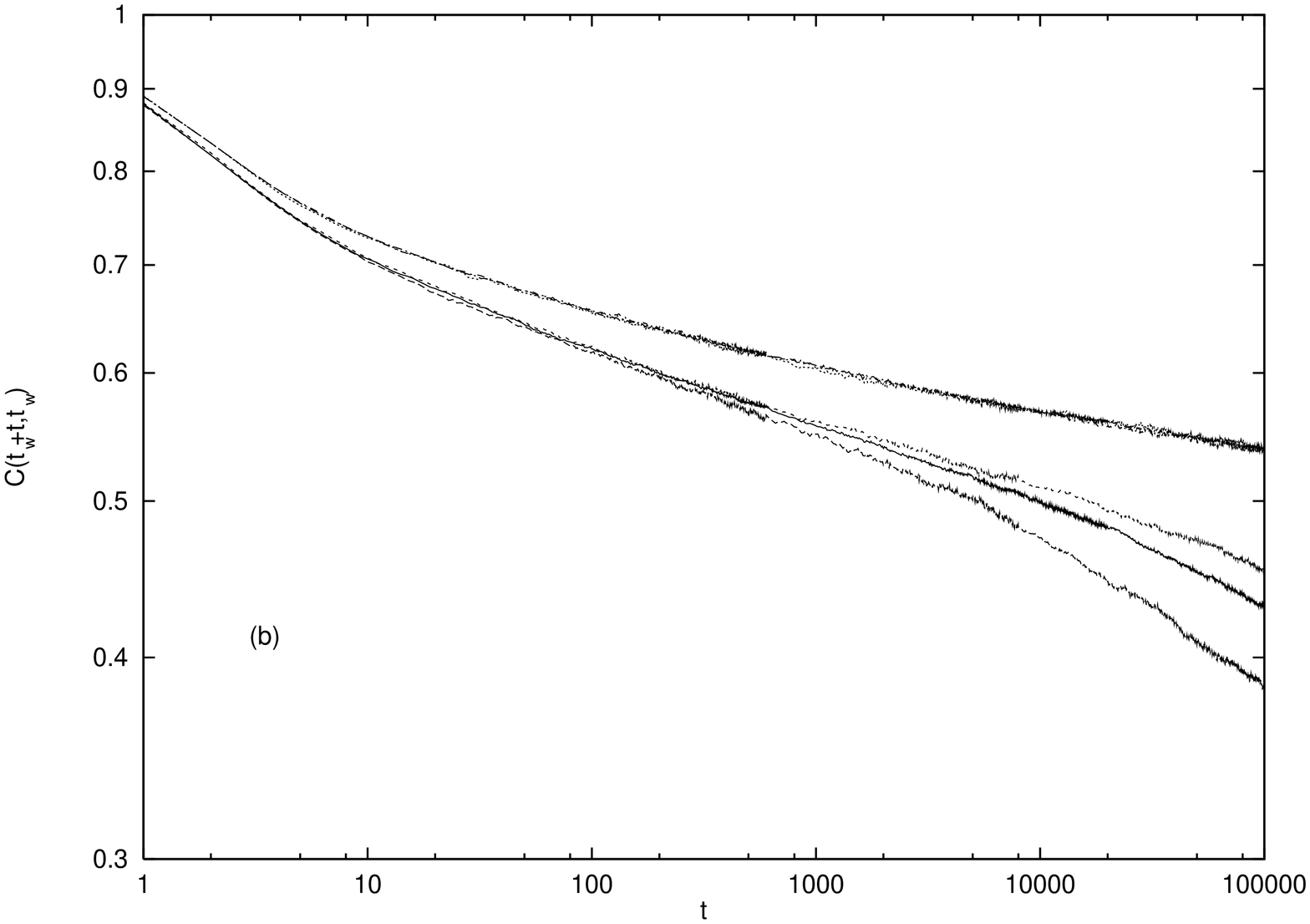,width=8.5cm,angle=-90}}
\caption{$C(t_w+t,t_w)$  versus $t$, in logarithmic scales~: comparison
of the evolutions for various initial conditions. For each
f\/igure, the three lower curves correspond to random initial conditions
((a), for $T=1$, $t_w=500,\ 5000,\ 100000$, and (b), for
$T=2$, $t_w=10000,\ 50000,\ 100000$), while three curves
corresponding to the system being prepared in its ground state
at initial time are superimposed onto each other
(with the same waiting times $t_w$ as for the lower curves),
showing that $C(t_w+t,t_w)$ depends only on $t$ in this case.
}
\label{cor2}
\end{figure}

The measure of the correlation function $C(t_w+t,t_w)$ for various
$t_w$ and $t$ shows in this case a simple $t$ dependence~: the system
is time translation invariant (f\/igure (\ref{cor2}).
Besides, $C(t_w+t,t_w)$ does not seem to go to zero for large
$t$, or at least stays well above the correlation at same times,
for a system with random initial condition,
for the accessible times~: no power law decay $t^{-\lambda}$
is found. It seems that ergodicity is really, and not simply
weakly, broken.

We have also studied the dynamics for an initial
conf\/iguration of energy close to the ground state energy, but
spatially well separated. In this case, we observe a similar
behaviour, with a time translation invariant correlation function.

Such behaviour has been observed
for example in the TAP states
of the $p$-spin spherical model with $p \ge 3$
\cite{babumezpspin}.

\section{Energy}

We have monitored the evolution of the energy density
$E_{al}$ of the polymer~; for a random initial condition
the initial energy is high, so the initial behaviour is a fast
decay, followed by a much slower evolution.

When the polymer is prepared in its ground state on the contrary, its
energy $E_f$ grows quickly because of the thermal bath, and then stays
constant.

\begin{figure}
\centerline{\epsfig{figure=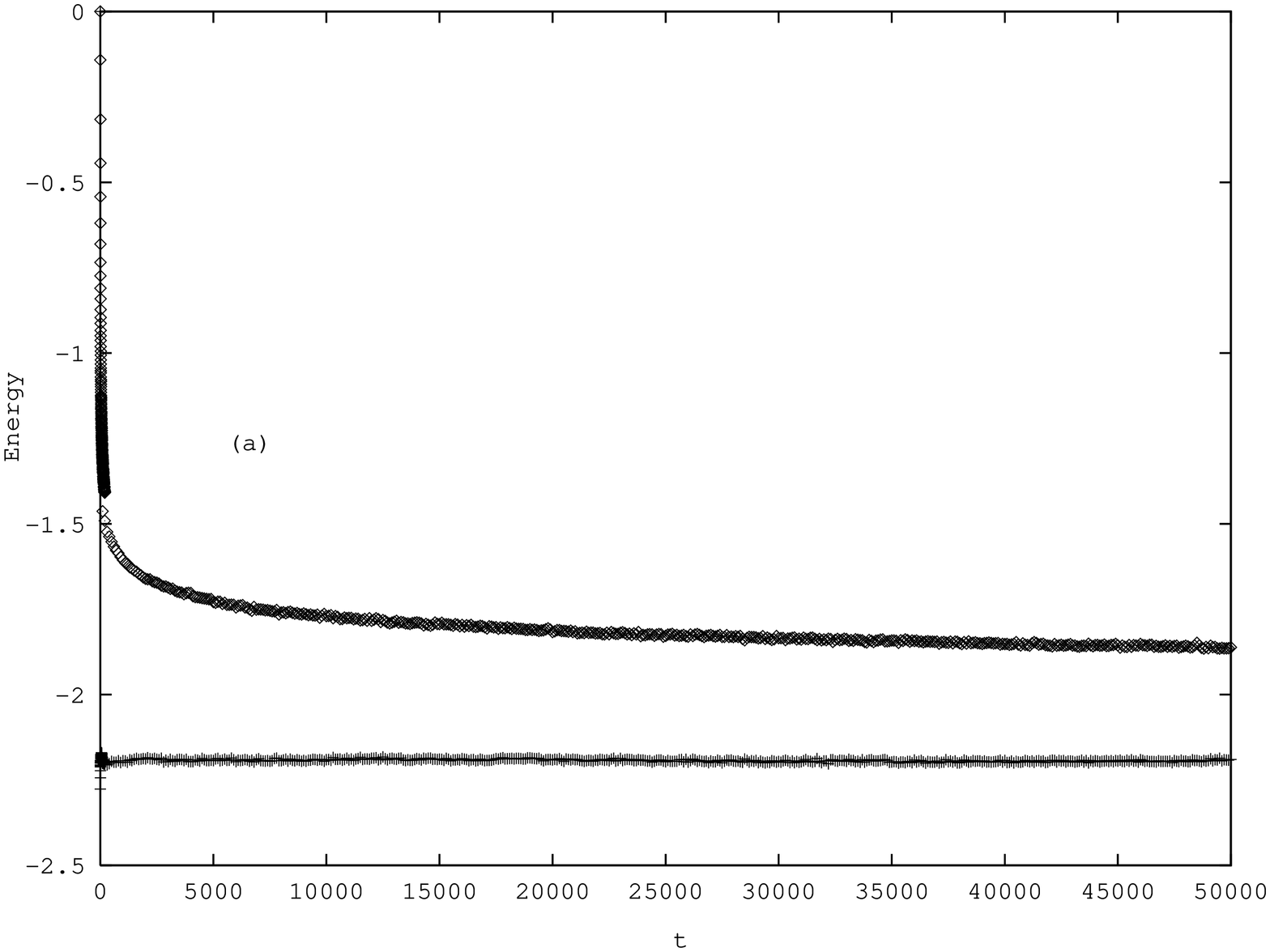,width=7.5cm,angle=-90}}
\vskip .8cm
\centerline{\epsfig{figure=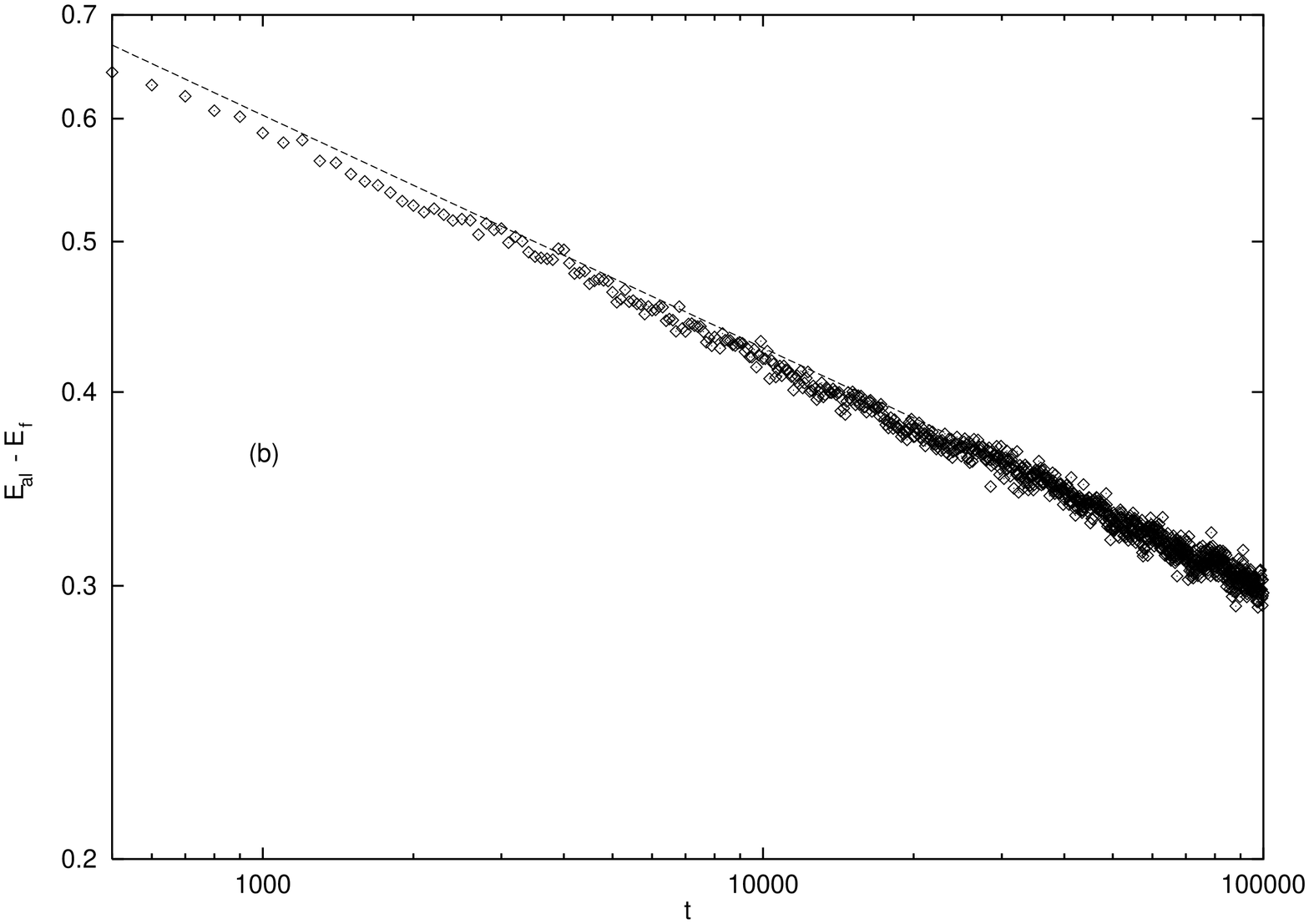,width=7.5cm,angle=-90}}
\caption{(a)~: evolution of the energies as a function of time, for
$T=2$, for random initial conditions (upper curve), or for a system
prepared in its ground state (lower curve).
(b)~: $E_{al}-E_f$ versus time, for $T=2$,
in logarithmic scale (symbols)~;
the straight line  corresponds to the power law $t^{-.15}$.}
\label{energie}
\end{figure}

We show in f\/igure (\ref{energie}) the dif\/ference between the energy
densities for both situations, in logarithmic scales~:
this plot shows that the evolution is compatible with a power law
decay of $E_{al}$ towards $E_f$. All the aging dynamics take
therefore place at higher energy densities than those of the
low lying states, where the dynamics is stationary.

\section{Discussion}

Whereas the previous analysis of the dynamics made by Yoshino
was based on an analogy with the droplets model
\cite{fisherhuse1,fisherhuse2},
we focus here more on a phase space analysis.

It must however be clear that, as for many numerical simulations,
especially for glassy systems, the available time scales remain
quite small, and that the results should therefore be
considered as tendencies, indications of behaviour. They allow
us, with these precautions, to present the following analysis.

The observed behaviour is quite similar to the one found for
the spherical $p$-spin model with $p=2$~: the dynamic consists
in a slow search of the ground state, with a slowly decaying
energy. There exists many states with low energy, but the polymer
is not able to f\/ind them, during its aging dynamics~: it remains
at higher energy density~; on the
contrary, if it is put in one of these conf\/igurations, it
stays trapped and has a stationary dynamics.

The behaviour of the overlap of two copies of the system, having
same conf\/igurations until a certain time, and then decoupled, shows
besides that the evolution takes place in some kind of ``gutters''
in phase space. The results of temperature cycling experiments also
show that the directed polymer in $1+1$ dimensions is a much simpler
system than real spin glasses. These two results are probably
related, and also in agreement, concerning the relative
simplicity of the phase space, with the fact
\cite{yoshino} that the response to a tilt field
applied at the end of the polymer does not display any aging~; however
it would be nice to measure the response to a spatially
sinusoidal field, and the relaxation after cutting the field,
to check if such a relaxation has aging, and how it depends
from the wavelength of the field. However, the
sample to sample fluctuations of
the response functions are very important, and such measurements are
therefore very difficult.

These behaviours are in fact intermediate between the $p=2$ spherical
$p$-spin model (analogous to domain growth \cite{cugldean})
and the $p \ge 3$ case which,
despite being a mean f\/ield model, displays a much more complicated
behaviour with long term memory and a very complex phase space
\cite{cukuprl}.

It would certainly be very interesting to study the directed polymer
dynamics in higher dimensions~: these new dimensions could provide a way
to avoid energy barriers by going around them. The appearance
of these entropy barriers \cite{barratmezard,felix,franzritort1,franzritort2}
could yield new interesting ef\/fects and a richer dynamic in a more complex
phase space.

It is a pleasure to thank M. M\'ezard who initiated this work
and R. Monasson for useful discussions and comments.


\section*{References}

\begin{thebibliography}{10}

\bibitem{revuepol}
T.~Halpin-Healy and Y.-C. Zhang,
\newblock Phys. Rep. {\bf 254}, 215 (1995).

\bibitem{interf}
D.~A. Huse and C.~L. Henley,
\newblock Phys. Rev. Lett. {\bf 54}, 2708 (1985).

\bibitem{quantum}
M.~Kardar,
\newblock Nucl. Phys. B {\bf 290}, 582 (1987).

\bibitem{revuevortex}
G.~Blatter, M.~V. Feigel'man, V.~B. Geshkenbein, A.~I. Larkin, and V.~M.
  Vinokur,
\newblock Rev. Mod. Phys. {\bf 66}, 1125 (1994).

\bibitem{derridaspohn}
B.~Derrida and H.~Spohn,
\newblock J. Stat. Phys. {\bf 51}, 817 (1988).

\bibitem{rem}
B.~Derrida,
\newblock Phys. Rev. Lett. {\bf 45}, 79 (1980).

\bibitem{imbrie}
J.~Z. Imbrie and T.~Spencer,
\newblock J. Stat. Phys. {\bf 52}, 609 (1988).

\bibitem{cook}
J.~Cook and B.~Derrida,
\newblock J. Stat. Phys. {\bf 57}, 89 (1989).

\bibitem{derridapol}
B.~Derrida and O.~Golinelli,
\newblock Phys. Rev. A {\bf 41}, 4160 (1990).

\bibitem{marc90}
M.~M\'{e}zard,
\newblock J. Phys. France {\bf 51}, 1831 (1990).

\bibitem{yoshino}
H.~Yoshino,
\newblock J. Phys. A {\bf 29}, 1421 (1996).

\bibitem{rieger}
H.~Rieger,
\newblock Annual Reviews of Computational Physics {\bf II}, 295 (1995).

\bibitem{fisherhuse1}
D.~S. Fisher and D.~A. Huse,
\newblock Phys. Rev. Lett. {\bf 56}, 1601 (1986).

\bibitem{fisherhuse2}
D.~S. Fisher and D.~A. Huse,
\newblock Phys. Rev. B {\bf 38}, 373 (1988).

\bibitem{hwafisher}
T.~Hwa and D.~S. Fisher,
\newblock Phys. Rev. B {\bf 49}, 3136 (1994).

\bibitem{cukuprl}
L.~F. Cugliandolo and J.~Kurchan,
\newblock Phys. Rev. Lett. {\bf 71}, 173 (1993).

\bibitem{franzmezard}
S.~Franz and M.~M\'ezard,
\newblock Physica A {\bf 210}, 48 (1994).

\bibitem{cuglledou}
L.~F. Cugliandolo and P.~L. Doussal,
\newblock Phys. Rev. E {\bf 53}, 1525 (1996).

\bibitem{cukuledou}
L.~F. Cugliandolo, J.~Kurchan, and P.~L. Doussal,
\newblock Phys. Rev. Lett. {\bf 76}, 2390 (1996).

\bibitem{mezpar91}
M.~M\'{e}zard and G.~Parisi,
\newblock J. Physique I {\bf 1}, 809 (1991).

\bibitem{andrea1}
A.~Baldassarri,
\newblock {\em Non-equilibrium Monte Carlo dynamics of the Sherrington
  Kirkpatrick mean field spin glass},
\newblock condmat 9607162  (1996).

\bibitem{andrea2}
A.~Baldassarri,
\newblock Studia della dinamica fuori dall'equilibrio di un modello di vetro di
  spin in campo medio,
\newblock Master's thesis, Universit\'a degli studi di Roma ``La Sapienza'',
  1995.

\bibitem{cugldean}
L.~F. Cugliandolo and D.~S. Dean,
\newblock J. Phys. A {\bf 28}, 4213 (1995).

\bibitem{babumezov}
A.~Barrat, R.~Burioni, and M.~M\'ezard,
\newblock J. Phys. A {\bf 29}, 1311 (1996).

\bibitem{saclay}
E.~Vincent, J.~Hammann, M.~Ocio, J.~P. Bouchaud, and L.~F. Cugliandolo,
\newblock {\em Slow dynamics and aging in spin glasses},
\newblock preprint condmat 9607224  (1996).

\bibitem{bouchaud}
J.~P. Bouchaud,
\newblock J. Phys. I France {\bf 2}, 1705 (1992).

\bibitem{riegercycling}
H. Rieger
\newblock J. Phys. I France {\bf 4}, 883 (1994).

\bibitem{babumezpspin}
A.~Barrat, R.~Burioni, and M.~M\'ezard,
\newblock J. Phys. A {\bf 29}, L81 (1996).

\bibitem{barratmezard}
A.~Barrat and M.~M\'ezard,
\newblock J. Phys. I (France) {\bf 5}, 941 (1995).

\bibitem{felix}
F.~Ritort,
\newblock Phys. Rev. Lett. {\bf 75}, 1190 (1995).

\bibitem{franzritort1}
S.~Franz and F.~Ritort,
\newblock Europhys. Lett. {\bf 31}, 507 (1996).

\bibitem{franzritort2}
S.~Franz and F.~Ritort,
\newblock condmat 9508133, to appear in J. Stat. Phys.  (1996).

\end{thebibliography}
\end{document}